\documentclass[aps,pra,preprint,groupedaddress]{revtex4}

\usepackage{graphicx}
\usepackage{dcolumn}
\usepackage{bm}
\usepackage{color}
\usepackage{amsfonts,amsmath,amsthm,graphicx}

\begin{document}

\title{Separating the output ports of a Bragg interferometer via velocity selective transport}

\author{R.~Piccon}
\author{S.~Sarkar}
\author{S.~Merlet}
\author{F.~Pereira~Dos~Santos}

\email{franck.pereira@obspm.fr}

\affiliation{LNE-SYRTE, Observatoire de Paris, Université PSL, CNRS, Sorbonne Université, 61 Avenue de l’Observatoire, 75014 Paris, France}

\date{\today}

\begin{abstract}
 We report on the study of a detection scheme based on a Bloch separator in two-photon Bragg interferometers. We increase the spatial separation between the two output ports of the interferometer by selectively imparting 30 Bloch oscillations to one of them before their detection via time of flight. This method allows increasing the duration of the interferometer by reducing the time for discriminating the ports of the interferometer at detection. We study in detail the impact of this separator on the performance of a dual gravity sensor, and in particular on its measurement sensitivities to gravity acceleration and gravity gradients.\\
\end{abstract}


\maketitle

\section*{Introduction}
Cold atom interferometers have demonstrated their ability to perform state of the art measurements of inertial quantities \cite{Geiger2020} such as gravity acceleration \cite{Louchet-Chauvet2011, Freier2016, Hu2013, karcher2018}, gravity gradient \cite{MCGuirk2002, Asembaum2017} or rotation \cite{Durfee2006,Savoie2018}, allowing to realize tests of fundamental physics, such as the test of the equivalence principle \cite{Asenbaum2020}, and the measurement of the fine structure constant \cite{Morel2020} or the gravitational constant G \cite{Fixlerg2007, Rosi2014}. In these devices, the inertial quantities of interest are extracted from phase measurements. This requires in general to measure the populations in the interferometer's output ports, thus demanding a clear separation between them. In the case of Raman interferometers, state labeling \cite{Borde1989} allows for detecting them separately with a series of properly tuned laser pulses, such as in the time of flight (TOF) detection method used in atomic fountain clocks \cite{Santarelli1999}. In the case of Bragg interferometers, the output ports are in the same hyperfine state and differ only by their momentum. One can thus in principle spatially resolve the different momenta states after some free evolution, via absorption imaging on a CCD camera for instance \cite{Gersemann2020}. This method ideally requires that the spatial separation between the output ports exceeds the size of the atomic cloud at the detection. While such a condition is readily fulfilled with Bose-Einstein condensates (BECs), owing to their small initial size and low expansion, or when using high-order Bragg diffraction \cite{D'amico2016, Muller2008}, this is generally not the case with laser cooled atomic samples undergoing low order Bragg diffraction processes. We explore a similar regime in this article, where the impact on the width of the TOF signals is more dominated by the convolution with the finite size of the detection than by the velocity-width of the atomic sample, thus leading to the requirement of a large free fall to separate the output ports at detection. Several detection methods have been developed so far to circumvent such limitations. One of them makes use of a velocity selective Raman pulse after the interferometer to transfer the desired output port to another hyperfine state and drive state-selective time of flight (TOF) detection \cite{Cheng2018}. Another method reported relies on delaying the interferometric pulses. This leads to resolved spatial fringes within the cloud and the possibility to retrieve out of them the phase of the interferometer \cite{Wigley2019}. 

As an alternative, we present here a method based on the use of a moving optical lattice to spatially separate the output ports of the interferometer before detecting them in a single resonant horizontal light sheet with different delays. This method was first used in \cite{Altin2013}, where a BEC was used as a source in an atom gravimeter. This was not found to induce any measurable noise in the data, the sensitivity actually being largely limited by vibration noise. In this article, we perform a comprehensive study of the impact of this detection method on the performances of a dual gravity sensor based on laser cooled atoms (rather than a BEC) that measures not only gravity acceleration but also gravity gradients.

\section*{Principle of the experiment}
The experimental setup, previously described in \cite{Caldani2019}, uses two vertically separated atom gravimeters to simultaneously measure gravity and its gradient. It is composed of two atomic sources of $^{87}Rb$ (top and bottom respectively) separated by 1 m. The distance from the center of the top source chamber to the top of the experiment is 45 cm, while the distance from the center of the bottom source chamber to the bottom of the experiment is 20 cm. Each source is based on a three dimensional (3D) surface magneto-optical trap (MOT) loaded from a 2D MOT. The overall height of the vacuum chamber is around 2 m and the detection region is located at the bottom of the experiment.

After a loading time of 680 ms, we collect around $10^{8}$ atoms in each source chamber, which we further cool down to $2~\mu K$ in a far detuned molasses. For cooling and detecting the atoms, we use two extended cavity diode lasers tuned on the $\vert F=2 \rangle \rightarrow \vert F'=3 \rangle$ cycling transition and $\vert F=1 \rangle \rightarrow \vert F'=2 \rangle$ repumping transition, amplified by a common tapered amplifier as in \cite{Merlet2014}. Atoms are then selected with a combination of Raman and pusher pulses in the $|F=1,m_{F}=0\rangle$ state with a narrower vertical velocity spread of $0.3~v_{r}$ rms, where $v_{r}$ is the recoil velocity of a photon. Once prepared, the two atomic clouds are launched upward simultaneously using a common accelerated lattice, coherently imparting Bloch oscillations to both clouds \cite{Langlois2017}. The final velocity of the atomic clouds corresponding to 110 Bloch oscillation is 1.3 m.s$^{-1}$. Atom interferometers are then driven simultaneously on the clouds using a sequence of $\frac{\pi}{2}$ - $\pi$ - $\frac{\pi}{2}$ two photon Bragg transitions, realized with Gaussian pulses of 28-56-28 $\mu$s full width at $1/e^{2}$ respectively. The Raman, Bragg and launch processes are realised with a single laser source described in \cite{Sarkar2022}. The maximum interrogation time used in this work is $2T=260$ ms where $T$ is the time separation between two consecutive Bragg pulses. We then finally detect the atoms by fluorescence.

\begin{figure}[ht]
    \centering
    \includegraphics[width=0.45\textwidth]{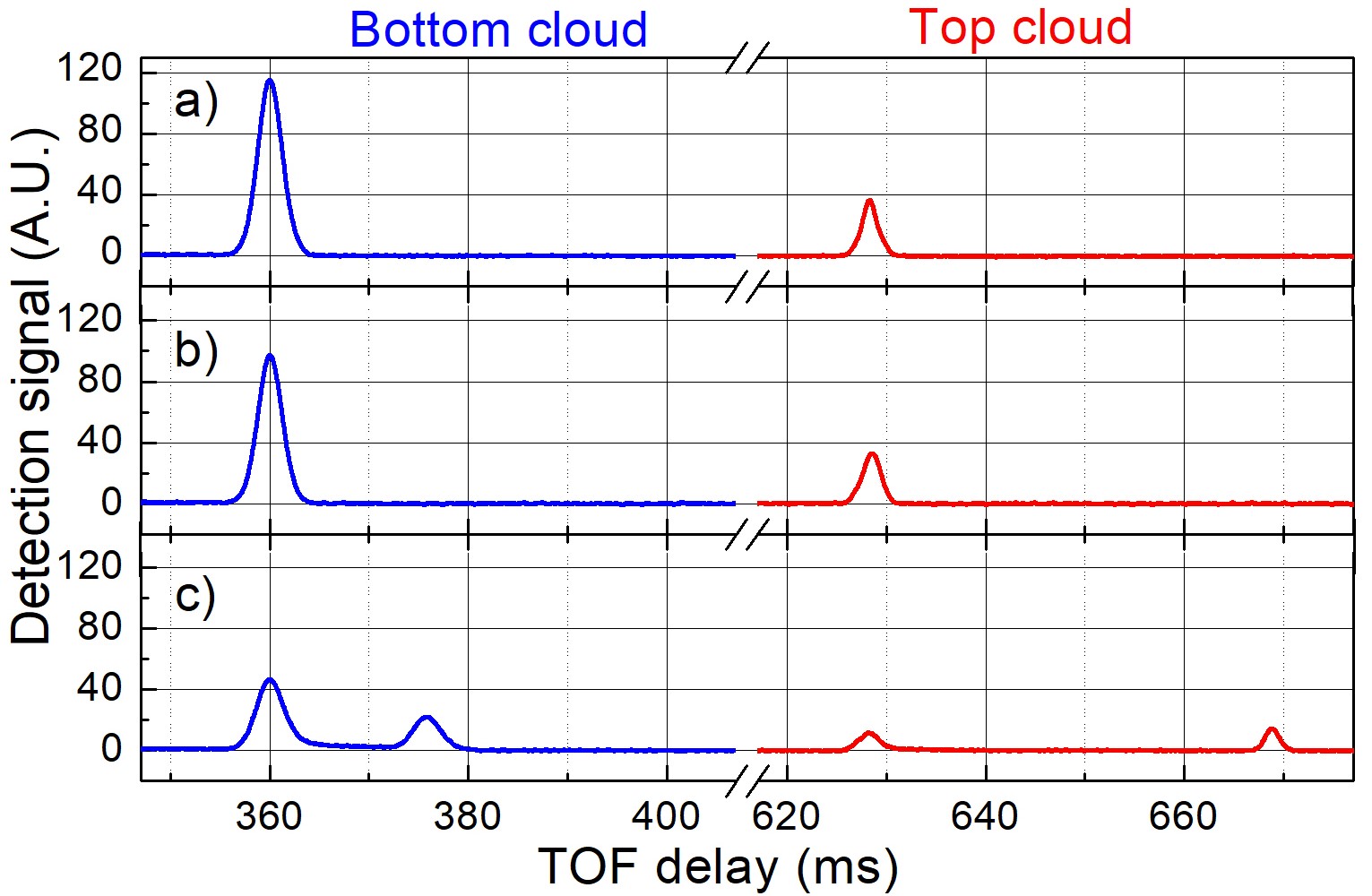}
    \caption{TOF signals of top and bottom atomic clouds. a) with neither Bragg pulse nor Bloch separator. b) with a Bragg $\pi/2$ pulse and no Bloch separator. c) with a Bragg $\pi/2$ pulse and a Bloch separator.}
    \label{toftrace}
\end{figure}

Our detection system is composed of three rectangular retro-reflected horizontal light sheets (15mm wide, 5mm high). The first one is tuned on the cyclic transition to detect the atoms in $\vert F=2 \rangle$ state and has its lower part not retro-reflected, so as to push the atoms afterwards. The second light sheet is tuned on the repumping transition to transfer the atoms in $\vert F=1 \rangle$ to $\vert F=2 \rangle$ before they fluoresce in the third, also tuned on the cyclic transition. This versatile setup allows for characterizing both Raman (used in our experiment for velocity selection) and Bragg processes. In particular, when the atoms are in the same hyperfine state, such as at the output of our Bragg interferometer, we obtain a maximal fluorescence signal by directing all the power of both detection and repumping beams into the first detection sheet only. A typical TOF signal for atoms launched upward is displayed in figure \ref{toftrace}a). The bottom (resp. top) cloud crosses the detection after a delay (as measured from the start of launch) of 360 ms (resp. 629 ms) with 2.45 ms (resp. 1.71 ms) full width at $1/e^{2}$.

\section*{The detection method}
In the work presented here, we use first order Bragg diffraction, which results in a momentum difference between the two output ports of $2 \hbar k$. Owing to the delays $\Delta t_{b}$ and $\Delta t_{t}$ between the last Bragg pulse and the arrival of the bottom and the top cloud at the TOF detection, the output ports of the two interferometers are spatially separated at the detection by distances of $\Delta z_{b}= \frac{2\hbar k}{m} \Delta t_{b}$ and $\Delta z_{t}= \frac{2\hbar k}{m} \Delta t_{t}$. Given our parameters, and in particular for an interferometer duration of $2T=260$ ms, $\Delta t_{b}= 87$ ms and $\Delta t_{t}= 356$ ms, which leads to $\Delta z_{b}=1.0$ mm and $\Delta z_{t}=4.2$ mm. This is comparable to, if not smaller than the size of the light sheet, thus preventing us from resolving these output ports. This is illustrated in figure \ref{toftrace}b), which displays the TOF signals obtained with clouds diffracted by a single $\pi/2$ Bragg pulse at the very time of the last interferometer pulse, which leads to equivalent momentum and spatial separations. A close look to the signals actually shows a slight widening and shift to the right, rather than a clear separation.

In order to increase the spatial separations $\Delta z_{b}$ and $\Delta z_{t}$ between the output ports, one solution could be to increase $\Delta t_{b}$ and $\Delta t_{t}$ by performing the last interferometer pulse earlier and reducing the interrogation time $T$ of the interferometer. But since the sensitivity of the gradiometer increases with $T$, the later the last Bragg pulse, the better. Alternatively, the use of an accelerated lattice circumvents this problem \cite{Altin2013}, allowing to induce a velocity selective transport of one of the output ports and thus to increase the spatial separation between them at the detection. We will name this process a Bloch separator throughout the article. In practice, the atoms are loaded adiabatically in the lattice for $250~\mu$s, before undergoing 30 Bloch oscillations in 5 ms, and being finally released adiabatically for another $250~\mu$s. The frequency difference between the two lattice beams is adjusted so as to slow down selectively the $2\hbar k$ output port of the interferometer. Atoms in this port thus get detected later, as displayed in figure \ref{toftrace}c), the time separation between the signal of the two output ports being now increased up to 15 ms for the bottom cloud and 42 ms for the top cloud on the detection signal. For each cloud, the TOF signal is finally adjusted with a double Gaussian function and the transition probability $P=\frac{N_2}{N_0+N_2}$ is calculated with $N_0$ (resp. $N_2$) the number of atoms in the $0\hbar k$ (resp. $2\hbar k$) output port. 

This technique reduces the separation time between the last interferometer pulse and the detection, thus maximizing the available interrogation time and the sensitivity of the gradiometer. On the other hand, given the finite momentum width of the atoms and the small momentum separation of $2\hbar k$ between the ports, significant cross couplings are present, since the efficiency of the transport of $2\hbar k$ atoms is not perfect and part of the $0\hbar k$ output port is also transported despite being out of resonance. The depth and the acceleration of the lattice must then be optimized in order to minimize these cross-couplings which have a direct impact on the effective contrast of the interferometer.

\subsubsection*{Optimisation of the lattice depth}

\begin{figure}[ht]
    \centering
    \includegraphics[width=0.45\textwidth]{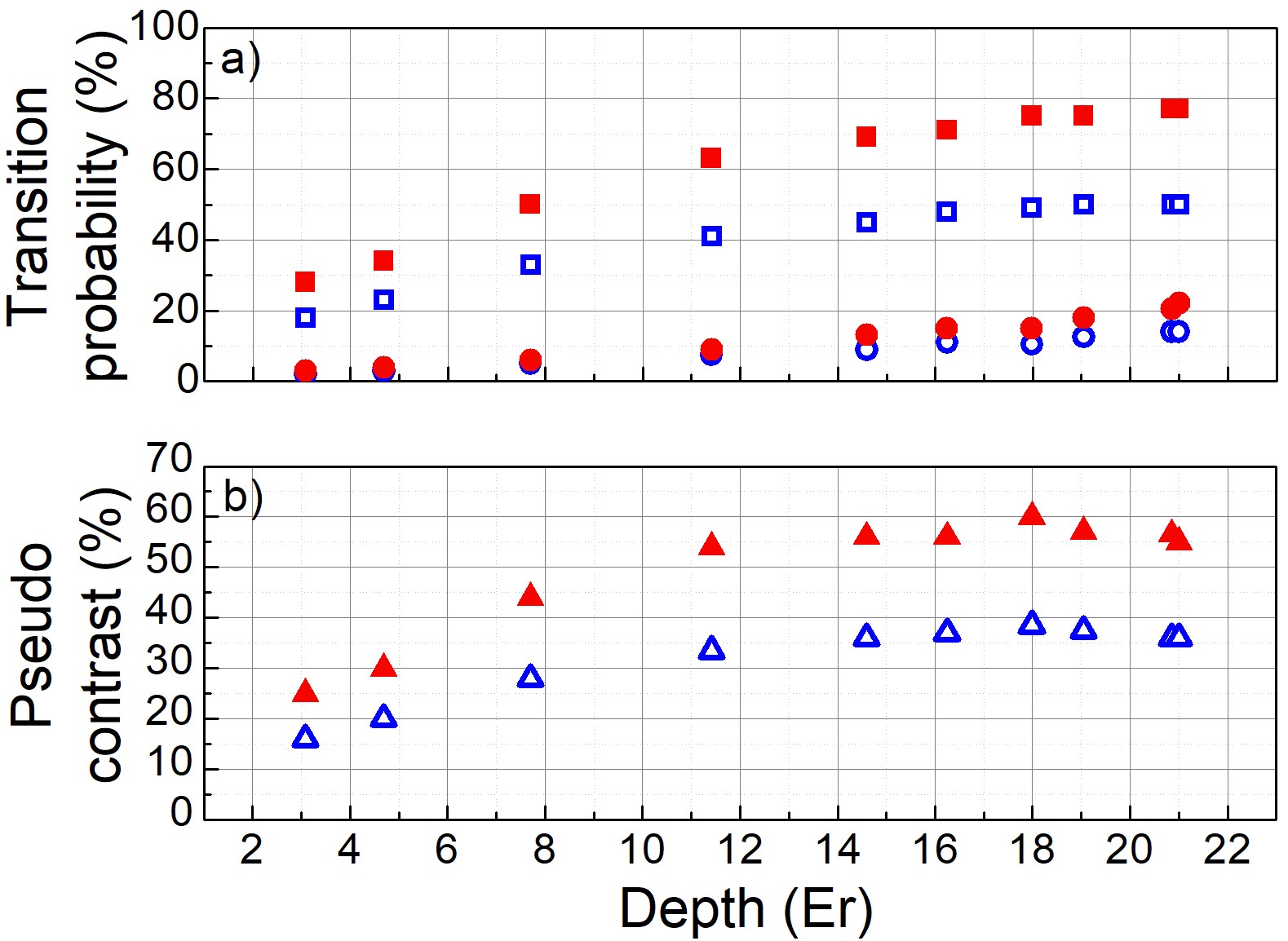}
    \caption{a) Transport efficiencies for the bottom (open blue symbols) and top (full red symbols) clouds versus lattice depth. Squares correspond to the transport efficiency of the target output port, while circles correspond to the parasitic transport of the unwanted output port. b) Pseudo contrast versus lattice depth. Bottom cloud: open blue triangles. Top cloud: full red triangles.}
    \label{contrast vs depth joint}
\end{figure}

Figure \ref{contrast vs depth joint}a) displays the transport efficiencies of the atomic clouds, after the velocity selection phase, as a function of the lattice depth. Square points (blue open for the bottom and red full for the top) are obtained by setting the frequency difference between the two lasers close to the resonance corresponding to the mean velocity of the cloud. Circles are obtained by shifting this frequency difference off resonance by 30 kHz, corresponding to a Doppler shift of $2\hbar k$, in order to evaluate what the parasitic transport efficiency of the unwanted output port will be. Note that these two frequency differences are actually shifted positively by 7.5 kHz with respect to resonance in order to avoid even larger cross-couplings due to non adiabatic loading of atoms in the upper Brillouin zone \cite{Clade2010}. The duration of the Bloch separator has been optimized for each lattice depth. The pseudo contrast displayed in figure $\ref{contrast vs depth joint}$b) is defined by the difference between the transport efficiencies close to resonance and off resonance of figure \ref{contrast vs depth joint}a). It characterizes the separation efficiency of the two output ports by the Bloch separator. An optimum of $60\%$ for the top cloud and $38\%$ for the bottom cloud is found for a depth of about 18 Er. These numbers actually represent the maximum effective contrasts that the detection scheme allows for. We attribute the difference in transport efficiencies and pseudo contrasts amplitude between the two clouds to different detection efficiencies of the atoms lost in the transport process, due to Landau-Zener tunneling and spontaneous emission, since the free fall times between the Bloch separator and the detection are different for the two clouds. 

\begin{figure}[ht]
\centering
\includegraphics[width=0.45\textwidth]{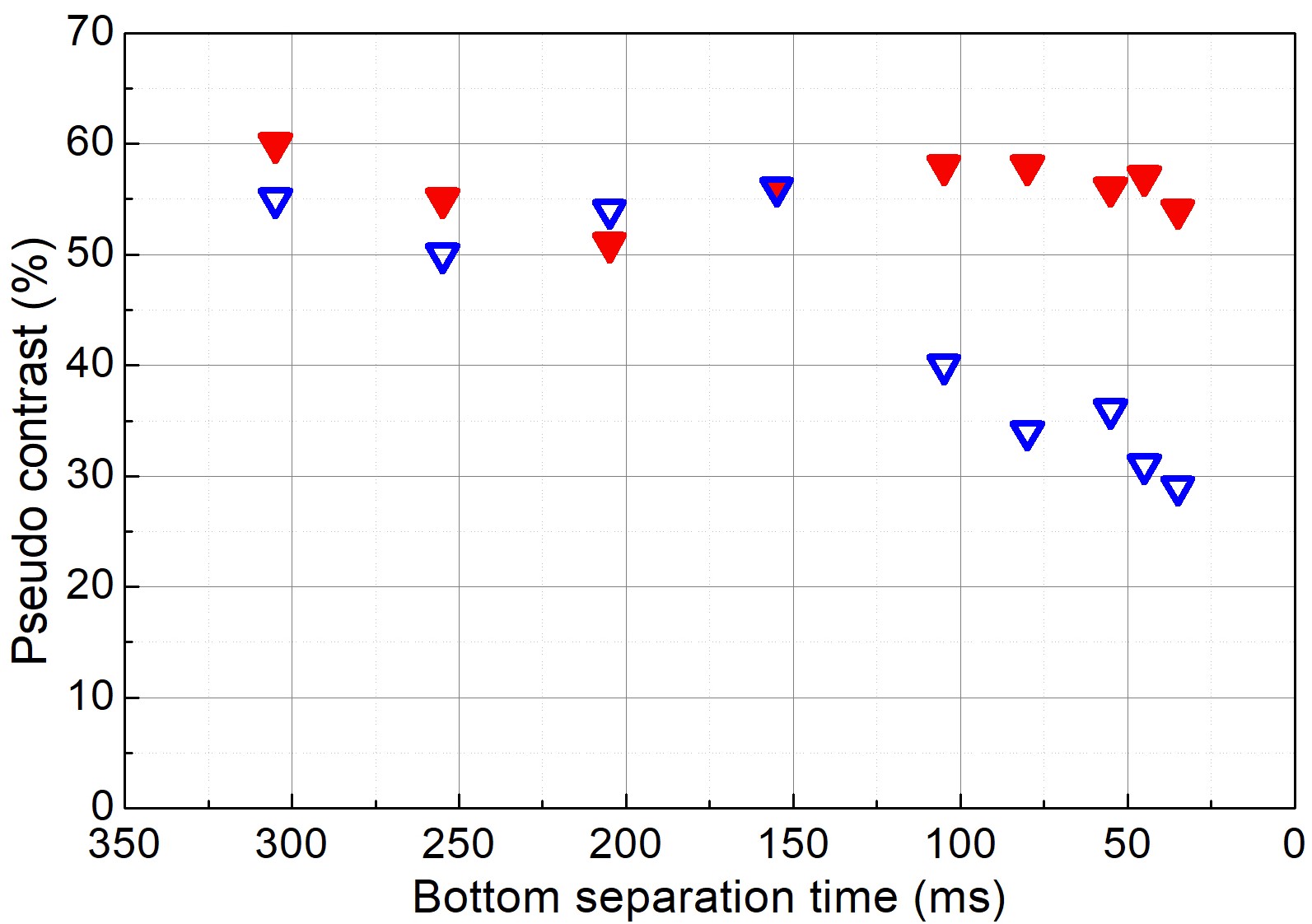}
\caption{Pseudo contrast (blue for the bottom and red for the top) as a function of the bottom separation time for a lattice depth of 18 Er.}
\label{contrastdelay}
\end{figure}

Finally, pseudo contrasts at the depth of 18 Er were measured as a function of the delay at which the Bloch separator is applied. The results are displayed on figure \ref{contrastdelay}, where the bottom separation time is defined as the delay between the Bloch separator and the detection of the bottom cloud. We observe a fairly constant pseudo contrast of $50-60\%$ for the top cloud. By contrast, increasing the bottom separation time clearly increases the pseudo contrast of the bottom cloud. A minimal delay of 150 ms is found to be necessary to obtain pseudo contrasts similar to the top cloud.

\subsubsection*{Optimisation of the bottom separation time}
\begin{figure}[ht]
    \centering
    \includegraphics[width=0.45\textwidth]{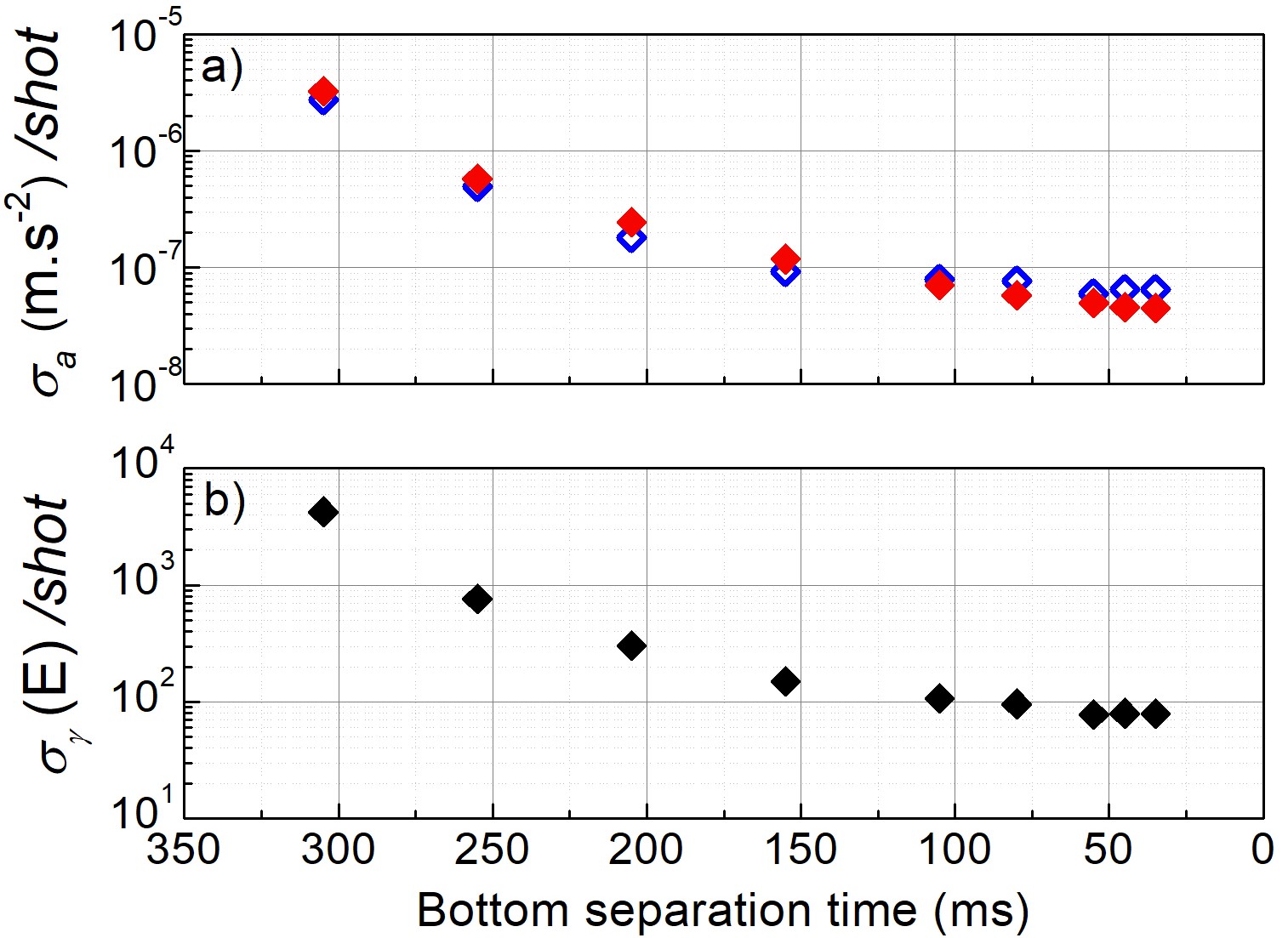}
    \caption{
    a) Sensitivities to gravity acceleration as a function of the bottom separation time. Bottom cloud: open blue diamonds. Top cloud: full red diamonds. b) Sensitivity to the vertical gravity gradient as a function of the bottom separation time. Measurements are performed at a lattice depth of 18 Er.}
    \label{sigma}
\end{figure}

The transition probability $P$ in the $2\hbar k$ output port is given by $P=A+\frac{C}{2}\cos(\Delta\phi)$ where $A$ is an offset, $C$ is the interferometer contrast and $\Delta\phi$ is the interferometer phase. For a gravimeter, $\Delta\phi=k_{eff}aT^{2}$, where $k_{eff}$ is the effective wave vector of the two-photon transition and $a$ is the acceleration. Assuming operation of the interferometer at mid-fringe, fluctuations of the phase are proportional to the fluctuations of the transition probability and the gravimetric sensitivity is given by: 

\begin{equation} \label{eq:1}
    \sigma_{a}=\frac{2\sigma_{P}}{CT^{2}k_{eff}}
\end{equation}

where $\sigma_{P}$ is the Allan standard deviation of the transition probability $P$. We now look for the optimum delay that maximizes the sensitivity $\sigma_{a}$. A late separator requires a larger number of Bloch oscillations to maintain the spatial separation between the output ports and thus increases losses and reduces the pseudo-contrast. An early separator reduces the available time for the interferometer and thus its scale factor and sensitivity. Figure \ref{sigma}a) displays the sensitivity to gravity acceleration as a function of the bottom separation time. This sensitivity is inferred using equation \ref{eq:1} with $2T$ the maximum interrogation time allowed by the separator, $C$ the pseudo-contrast given by figure \ref{contrastdelay} and measured levels of detection noise ($\sigma_{P}=3.5\times 10^{-3}$ for the bottom cloud and  $4.5\times 10^{-3}$ for the top one). Note that the sensitivities are given per shot, the cycle time being $T_{C}=1.45$ s.

Figure $\ref{sigma}$a) shows that the bottom cloud sensitivity ($\sigma_{a,b}$) improves when reducing the bottom separation time until reaching a plateau of $6\times10^{-8}$ m.s$^{-2}$/shot for bottom separation times below $55$ ms, for which the gain in interrogation time tends to be balanced by the loss of contrast. On the contrary, the top cloud sensitivity ($\sigma_{a,t}$) improves continuously.\\
Assuming uncorrelated detection noise for the two clouds, the sensitivity to the gravity gradient $\sigma_{\gamma}$ can be expressed as follow : $\sigma_{\gamma}=\frac{\sigma_{\delta a}}{L}=\frac{1}{L}\sqrt{\sigma_{a,b}^{2}+\sigma_{a,t}^{2}}$. Figure \ref{sigma}b) displays the calculated sensitivity to gravity acceleration and shows an optimum of about $80$ E/shot for a bottom separation time smaller than 55 ms.

\begin{figure}[ht]
    \centering
    \includegraphics[scale=0.27]{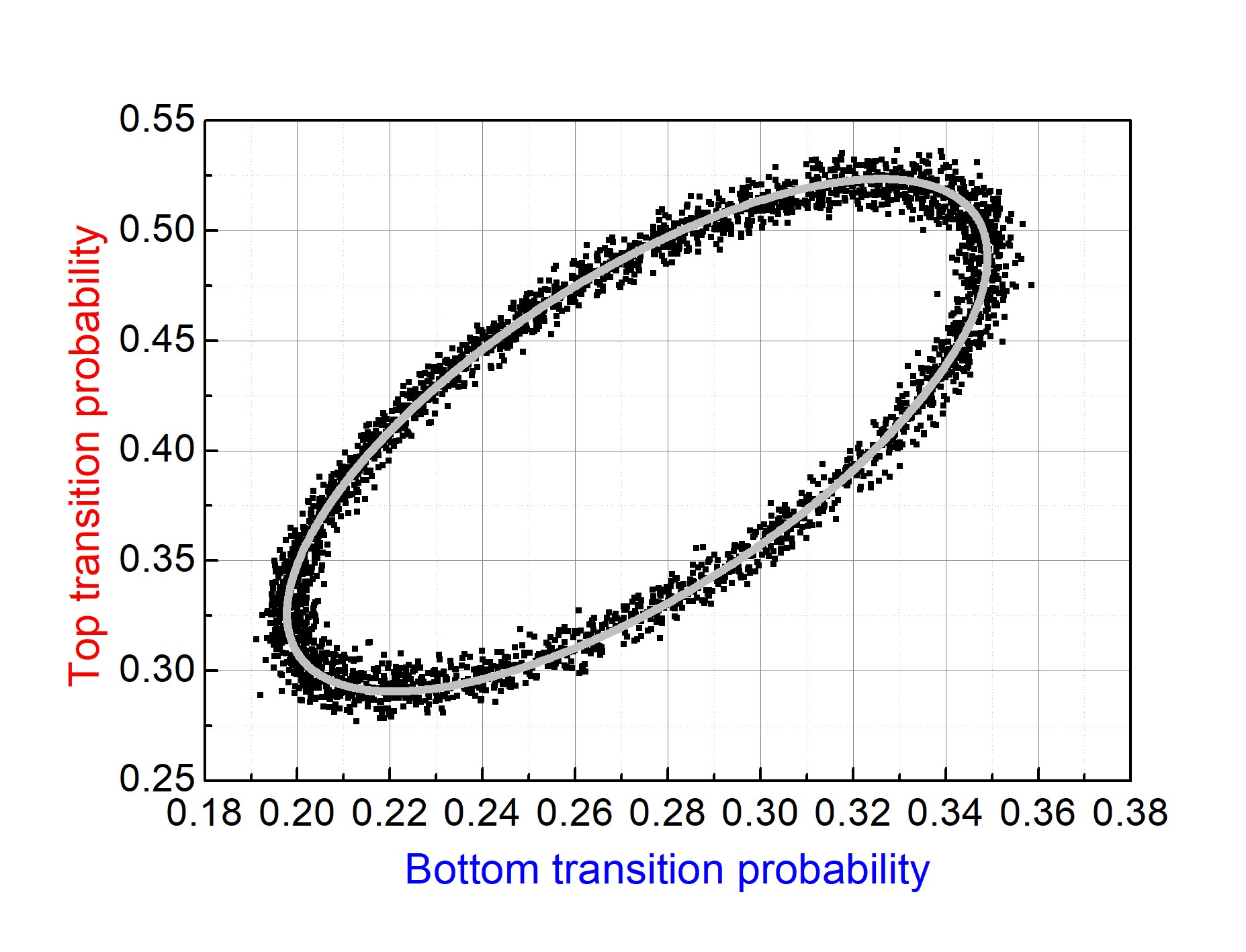}
    \caption{Parametric plot of the top and bottom transition probabilities for an interferometer duration 2T of 260 ms. Black dots: experimental data, grey line: ellipse fit to the data.}
    \label{ellipse}
\end{figure}

\section*{Gravity gradient measurement}
With optimized Bloch separator parameters (duration of 5ms, lattice depth of 18 Er, bottom separation time of 55 ms), we finally perform differential interferometer measurements. The ellipse resulting from the parametric plot of the two transition probabilities is represented in figure $\ref{ellipse}$, out of which the differential phase is extracted via ellipse fitting. We obtain an equivalent gradiometric sensitivity of $\sigma_{\gamma}= 345$ E/shot, larger than the $80$ E/shot lower bound imposed by the detection method. This is due on one hand to actual contrasts being smaller than pseudo contrasts because of Bragg pulses imperfections (arising  from coupling inhomogeneities and dephasing coming from Coriolis accelerations). On the other hand, the setup not being isolated from ground vibrations, the interferometer phase spans over about $4 \pi$ radians. The interferometers thus do not operate at mid fringe, and the extraction of the differential phase via ellipse fitting reduces the sensitivity by a factor $\sim 2$ with respect to a mid-fringe lock method \cite{Caldani2019}.

\section*{Conclusion}
We have studied the efficiency of a Bloch separator, a method based on the use of velocity selective Bloch oscillations to increase the spatial separation between the output ports of Bragg interferometers, easing their detection and allowing for maximizing the interferometer duration. For an interferometer duration of $2T=260$ ms, we evaluate for our experiment optimal acceleration sensitivities of $6\times10^{-8}$ m.s$^{-2}$/shot and a sensitivity to gravity gradients of $80$ E/shot. Future improvements will consist in increasing $2T$ up to $500$ ms, isolating the experiment from ground vibrations thanks to a passive isolation platform, compensating Coriolis acceleration thanks to a tip-tilt mirror and improving the Bragg laser beams intensity profiles with a flat top collimator. This should allow to reach state of the art gradiometric sensitivity, in the range of a few tens of Eötvös at one second \cite{MCGuirk2002, Asembaum2017}.\\

\section*{Acknowledgement}
The authors would like to acknowledge the financial support by CNES (R\&T R-S15/SU-0001-048 and R-S19/SU-0001-048), the CNRS program Gravitation, Références, Astronomie, Métrologie (PN-GRAM) and the ANR under contract ANR-19-CE47-0003 GRADUS. R. P. thanks CNES and DGA for financial support.

\end{document}